\pdfoutput=1

\RequirePackage{fix-cm}
\RequirePackage{rotating}

\documentclass[smallextended]{svjour3}
\smartqed 
\usepackage{cite}
\usepackage{mathptmx}
\usepackage{amsmath,amssymb,amsfonts}
\usepackage{algorithmic}
\usepackage{graphicx}
\usepackage{subcaption}
\usepackage{textcomp}
\usepackage{xcolor}
\usepackage{listings}
\usepackage{multirow}
\usepackage{wrapfig}
\usepackage{xcolor,colortbl}
\usepackage[flushleft]{threeparttable}
\usepackage{amsmath}
\usepackage{relsize}
\usepackage{pifont}
\usepackage[sort&compress,numbers]{natbib}
\usepackage[skins]{tcolorbox}
\usepackage{wasysym}
\usepackage{pgfplots}
\usepackage{changepage}

\usepackage{cite}

\newcommand{\bi}{\begin{itemize}}
\newcommand{\ei}{\end{itemize}}
\newcommand{\be}{\begin{enumerate}}
\newcommand{\ee}{\end{enumerate}}

\usepackage{subcaption,siunitx,booktabs}

\pgfplotsset{compat=1.10}
\usepgfplotslibrary{fillbetween}

\def\BibTeX{{\rm B\kern-.05em{\sc i\kern-.025em b}\kern-.08em
    T\kern-.1667em\lower.7ex\hbox{E}\kern-.125emX}}
    
\definecolor{MyDarkBlue}{rgb}{0,0.08,0.45} 
\newcommand\MyBox[2]{
  \fbox{\lower0.75cm
    \vbox to 1.7cm{\vfil
      \hbox to 1.7cm{\hfil\parbox{1.4cm}{#1\\#2}\hfil}
      \vfil}%
  }%
}

\begin{document}

\title{Omni: Automated Ensemble with Unexpected Models against Adversarial Evasion Attack
}

\author{Rui Shu         \and
        Tianpei Xia     \and
        Laurie Williams \and
        Tim Menzies
}

\institute{Rui Shu, Tianpei Xia, Laurie Williams, Tim Menzies\at
              Department of Computer Science, North Carolina State University, Raleigh, NC, USA \\
              Email: rshu@ncsu.edu, txia4@ncsu.edu, lawilli3@ncsu.edu, timm@ieee.org. \\We assert that the authors have no conflict of interests
}

\date{Received: date / Accepted: date}

\maketitle

\begin{abstract} {\small 
 \textbf{Background:} Machine learning-based security detection models have become prevalent in modern malware and intrusion detection systems. However, previous studies show that such models are susceptible to adversarial evasion attacks. In this type of attack, inputs (i.e., adversarial examples) are specially crafted by intelligent malicious adversaries, with the aim of being misclassified by existing state-of-the-art models (e.g., deep neural networks). Once the attackers can fool a classifier to think that a malicious input is actually benign, they can render a machine learning-based malware or intrusion detection system ineffective.

\textbf{Goal:} To help security practitioners and researchers build a more robust model against non-adaptive, white-box and non-targeted adversarial evasion attacks through the idea of ensemble model.

\textbf{Method:} We propose an approach called \textit{\textbf{Omni}}, the main idea of which is to explore methods that create an ensemble of ``unexpected models''; i.e., models whose control hyperparameters have a large distance to the  hyperparameters of an adversary's target model, with which we then make an optimized weighted ensemble prediction.

\textbf{Result:} In studies with five types of adversarial evasion attacks (FGSM, BIM, JSMA, DeepFool and Carlini-Wagner) on five security datasets (NSL-KDD, CIC-IDS-2017, CSE-CIC-IDS2018, CICAndMal2017 and the Contagio PDF dataset), we show
\textit{\textbf{Omni}} is a promising approach as a defense strategy against adversarial attacks when compared with other baseline treatments.

\textbf{Conclusion:} When employing ensemble defense against adversarial evasion attacks, we suggest to create ensemble with unexpected models that are distant from the attacker's expected model (i.e., target model) through methods such as hyperparameter optimization.

\textbf{Declarations:} \\
{\em Funding:} This work was partially funded by NSF grant \#1909516.\\
{\em Conflicts of interest/Competing interests:} The authors have no relevant financial or non-financial interests to disclose.

 }

\end{abstract}
 \newpage

\section{Introduction}

With the growing reliance on information technology, cybercrime is a serious threat to the economy, military and other industrial sectors~\cite{bissell2019cost,jang2014survey,chang2019evaluating,opderbeck2015cybersecurity,rovskot2020cybercrime}. 
In 2019, the damage cost caused by malware and cybercrime exceeded a trillion dollars~\cite{morgan2019official}. For example, a March 2019 ransomware attack on aluminum producer Norsk Hydro caused 60 million pounds of remediation cost~\cite{sechel2019comparative}. The attack brought production to a halt at 170 sites around the world. More generally, a 2019 study by Accenture reports that cybercrime will cost US \$5.2 trillion over the next five years~\cite{bissell2019cost} Alarming, that cost is growing.  That same report documents that in the United States, the annual average cost to organization of malicious software has grown 29\% in the last year. 

To counter those threats, machine learning algorithms are being widely applied
to security critical tasks, such as malware detection and intrusion detection. For example, security practitioners and researchers build detection models that utilize learned patterns to detect whether a new file (e.g., PDF file) or an application (e.g., Android app) or a network traffic becomes a security threat ~\cite{maiorca2013looking,smutz2012malicious,bhat2019survey,arp2014drebin}. But paradoxically, machine learning also introduces a new attack vector for motivated adversaries\cite{barreno2006can,DBLP:journals/pr/BiggioR18}. An active research field called \textit{adversarial machine learning} has received a significant amount of attention over the last decade. Adversarial machine learning is a technique that attempts to fool or misguide a machine learning-based model with malicious inputs. This technique was
first studied in spam filtering~\cite{dalvi2004adversarial,lowd2005adversarial,lowd2005good} and later on since 2014, Szegedy et al.~\cite{szegedy2013intriguing} found that small perturbations in images can cause misclassification in neural
network classifiers which attracted more studies in domains such as computer vision. While at the same time, this technique is also widely studied in the security domain. \textit{Adversarial evasion attack}~\cite{biggio2013evasion} is one of the most prevalent types of adversarial machine learning attacks that happens during the testing stage in the machine learning pipeline. In this attack, attackers try to evade the detection system by manipulating the testing data, resulting in a wrong model classification. The core of adversarial evasion attack is that when an attacker can fool a classifier to think that a malicious input (e.g., malicious Android application or network traffic) is actually benign, they can render a machine learning-based malware detector or intrusion detector ineffective.

Prior studies have tried~\cite{DBLP:conf/iclr/TramerKPGBM18,smutz2016tree,kantchelian2016evasion} to thwart evasion attacks by building a more robust and complex model via \textit{ensemble learning}~\cite{dietterich2002ensemble}. Ensemble learning is the process of (a)~building multiple models and then (b)~polling across the models to arrive at a final decision. In theory, ensemble learning tends to defend against adversarial evasion attack because attackers have to craft payloads (i.e., adversarial examples) that are able to subvert all constituent models at once, which makes it more difficult to be successful.
However, some other studies~\cite{zhang2020decision,zhang2018gradient,he2017adversarial,DBLP:conf/iclr/TramerKPGBM18,DBLP:journals/corr/PapernotMG16} caution that adversaries can still defeat ensemble-based strategy. For example, Papernot et al.~\cite{DBLP:journals/corr/PapernotMG16} find that, even when ensemble classifiers are used, adversarial attackers can still manage to use their own models to find ways to ``transfer'' the adversarial examples to a victim model (i.e., attacker's target model), even if (a)~the adversary has little knowledge of the victim model; and even if (b)~the defender uses an ensemble of classifiers.

The starting point for our research in this paper is the following observation. Prior work on ensemble learning against adversarial evasion attack~\cite{DBLP:conf/iclr/TramerKPGBM18,smutz2016tree,kantchelian2016evasion} barely explored the range of options available within a model or explored few different models. For example, some researchers build their ensemble-based approaches using a small number of constitute models; e.g., Kantchelian et al.~\cite{kantchelian2016evasion} use seven constitute models in their ensemble classifier. Such kind of small size ensemble classifier usually does not yield the optimal prediction performance~\cite{hernandez2013large}. However, as shown in Table~\ref{tbl:hyperparameters}, malware or intrusion detection models can be built from a space of trillions of options (i.e., hyperparameter choices of a model). The small space of ensembles used in previous work barely scratches the surface of the large space of options within ensemble generation.

\begin{table}[t]
\centering
\caption{Hyperparameters selected to tweak for deep neural networks and their ranges in hyperparameter optimization. Assuming the drop out rate is divided into 100 options, then this table shows a close to trillion options:
$11*11*120*20*22*100*7*4*15*3\approx 10^{13}$.}
\begin{threeparttable}
\begin{tabular}{c|l}
\hline
\rowcolor[HTML]{ECF4FF} \textbf{Hyperparameter} & \multicolumn{1}{c}{\cellcolor[HTML]{ECF4FF}\textbf{Ranges}} \\ \hline
\begin{tabular}[c]{@{}c@{}}Hidden layer\\ activation function\end{tabular} & \begin{tabular}[c]{@{}l@{}}elu, relu, selu, sigmoid, softmax, tanh, hard\_sigmoid, softplus,  \\ softsign,  linear, exponential\end{tabular} \\ \hline
\begin{tabular}[c]{@{}c@{}}Output layer\\ activation function\end{tabular} & \begin{tabular}[c]{@{}l@{}}elu, relu, selu, sigmoid, softmax, tanh, hard\_sigmoid, softplus, \\ softsign,  linear, exponential\end{tabular} \\ \hline
First layer dense & quniform(30, 150, 1) \\ \hline
Second layer dense & quniform(30, 50, 1) \\ \hline
Third layer dense & quniform(10, 32, 1) \\ \hline
Drop out rate & uniform(0.0, 0.5) \\ \hline
Optimizer & \begin{tabular}[c]{@{}l@{}}Adadelta, Adagrad, Adam, Adamax, \\ NAdam, RMSprop, SGD\end{tabular} \\ \hline
Batch size & 16, 32, 64, 128 \\ \hline
Number of epochs & quniform(5, 20, 1) \\ \hline
Learning rate & 0.001, 0.01, 0.1 \\ \hline
\end{tabular}
\begin{tablenotes}
\item * Note: \textbf{quniform($low$, $high$, $q$)} is a function returns a value like $round(uniform(low, high)/q) * q$, while \textbf{uniform($low$, $high$)} returns a value uniformly between $low$ and $high$.
\end{tablenotes}
\end{threeparttable}
\label{tbl:hyperparameters}
\end{table}

Accordingly, the core innovation of this paper is the proposed approach called \textit{\textbf{Omni}}. This method uses {\em hyperparameter optimization}~\cite{feurer2019hyperparameter} algorithms that build an ensemble system by exploring trillion of model options. Such optimization algorithms seek the set of hyperparameters of a given machine learning algorithm which return the optimal evaluation performance. In a machine learning algorithm, hyperparameters are properties that control the behaviors of the machine learning algorithms. For example, when reasoning about ``$k$-nearest neighbors'', the hyperparameter ``$k$'' decides how many neighbors to use for making a decision.

While exploring a large number of models (such as the trillions of options from Table~\ref{tbl:hyperparameters}), our \textit{\textbf{Omni}} method learns the ``expected model''; i.e. the optimal model from hyperparameter optimization, which is used in normal prediction. This model is the target of attackers and hence then becomes the victim model. Next, \textit{\textbf{Omni}}'s optimizer surveys the hyperparameter space of models to find ``unexpected models''; i.e. models that are (a)~performing well (i.e., sub-optimal models) and (b)~dissimilar to the expected model (i.e., in the architecture). More specifically:
\begin{itemize}
\item 
For hyperparameter optimization, we search a large space of possible configurations of a model to initialize a large \textit{model pool}.
\item
The {\em expected model} is a model from the model pool that performs best (under no attack). We call this model ``\textit{expected}'' since we conjecture that this model would be the target of an attacker, and becomes a victim model.
\item
We introduce an idea called \textit{model distance}, which is a numeric value indicating the degree of similarity of two models' hyperparameter configurations. A large model distance value means that two models are more likely to be different in their model architecture. 
\item The {\em unexpected models} are those whose performance within some small $\epsilon$ of the expected model but are more than some distance $t$ away from the expected model.   
\item
Those unexpected models are combined into a \textit{weighted ensemble}, in which each model $m_i$ in the ensemble with a weight $w_i$. The final prediction of this ensemble is the combined prediction of each model $m_i$ times its weight $w_i$. These weights are further optimized by an evolutionary algorithm that finds optimal $w_i$ setting that maximizes prediction performance.
\item
The final weighted ensemble, with its optimized weight setting is then deployed against evasion attacks.
\end{itemize}

The rest of the paper shows empirically that the results of using \textit{\textbf{Omni}} as a defense method against adversarial evasion attacks is promising. Background and related work is discussed in Section~\ref{sec:background}. We introduce threat model, adversarial evasion attack strategies as well as the proposed approach in Section~\ref{sec:method}. We then describe our datasets and experiment rigs in Section~\ref{sec:experiment}. For this studies, we use:
\begin{itemize}
\item
Five sophisticated adversarial evasion attack strategies; i.e. FGSM~\cite{DBLP:journals/corr/GoodfellowSS14}, BIM~\cite{KurakinGB17a}, JSMA~\cite{papernot2016limitations}, DeepFool~\cite{moosavi2016deepfool} and Carlini-Wagner~\cite{carlini2017towards,DBLP:conf/ccs/Carlini017};
\item
And five security datasets; i.e. NSL-KDD~\cite{nsl-kdd}, CIC-IDS-2017~\cite{sharafaldin2018toward}, CSE-CIC-IDS2018~\cite{sharafaldin2018toward}, CICAndMal2017~\cite{lashkari2018toward} and the Contagio PDF dataset~\cite{contagio-pdf}.
\end{itemize}
 
As shown in Section~\ref{sec:result} (i.e., the results section), \textit{\textbf{Omni}} demonstrates
its advantages when compared with other baseline treatments such as adversarial training, random ensemble and average weight ensemble. We discuss more about the nature of \textit{\textbf{Omni}} and other issues in Section~\ref{sec:discussion} and threats to validity in Section~\ref{sec:threats} then present the conclusion and future directions to extend this work in Section~\ref{sec:conclusion}. Here we conclude:
\begin{quote}
{\em
A well-designed weighted ensemble system is a promising approach to defend against adversarial evasion attack.
}\end{quote}
and
\begin{quote}
{\em
When using ensemble learning as a defense method against adversarial evasion attacks, we suggest to create ensemble with unexpected models that are distant from the attacker's expected model (i.e., target model) through methods such as hyperparameter optimization.
}\end{quote}

\newpage
\section{Background and Related Works}~\label{sec:background}

\subsection{Machine Learning for Security \& Adversarial Machine Learning}

The global security threat continues to evolve at a rapid pace, with a rising number of types of threats. For example, previous studies have shown many kinds of malicious software:

\begin{itemize}
\item
Malware that is a malicious file or hidden within files; e.g. PDF files can carry malicious code~\cite{smith01}.
\item
Ransomware that encrypts a victim's files then demands a ransom from the victims to restore access to the data upon payment. For example, a March 2019 ransomware attack on aluminum producer Norsk Hydro caused 60 million pounds of remediation cost. The attack brought production to a halt at 170 sites around the world (some 22,000 computers were affected across 40 different networks)~\cite{sechel2019comparative}. 
\item
Industrial espionage software that infects, then destroyed industrial machinery; e.g. the Stuxnet virus used to damage centrifuges at Iran’s Natanz uranium enrichment facility~\cite{langner2011stuxnet}.
\item
In the social network realm, Facebook estimated that hackers stole user information from nearly 30 million people through malicious software~\cite{facebookreport}.
\item
According to the International Data Corporation (IDC), the Android operating system covers around 85\% of the world’s smartphone market. Because of its increasing popularity, Android is drawing the attention of malware developers and cyber-attackers. Android malware families that are popular are spyware, bots, Adware, Potentially Unwanted Applications (PUA), Trojans, and Trojan spyware, which affect millions of Android users~\cite{bhat2019survey}.
\item
Other kinds of malicious software~\cite{monshizadeh2014security} include (a) scareware that socially engineers anxiety, or the perception of a threat, to manipulate users into e.g. buying unwanted software; (b) adware that throws advertisements up on your screen (most often within a web browser); and (c)~software that infects your computer then, without your permission of knowledge, mounts a denial of service attack on other computers. 
\end{itemize}

Security practitioners now routinely add security detectors to their environments, which are machine learning models that utilize known detective patterns to verify whether an application becomes a threat. Such detectors can be built in many ways including (but not restricted to) building a classifier to examine a web page for malicious content~\cite{canali2011prophiler,eshete2012binspect}; constructing multiple classifier systems to classify spam emails~\cite{biggio2010multiple}; building classifiers to detect malicious PDF files~\cite{xu2016automatically};
applying machine learning to detect Android malware~\cite{grosse2017adversarial}; designing supervised learning algorithm to classify HTTP logs~\cite{liu2017robust}; designing machine learning models to detect ransomware~\cite{munoz2017towards}; and detecting malicious PowerShell commands using deep neural networks~\cite{hendler2018detecting}.

Previous research works on adversarial machine learning are mainly focused on computer vision which solves tasks such as image recognition~\cite{DBLP:journals/corr/GoodfellowSS14,DBLP:conf/ccs/PapernotMGJCS17}. However, adversarial machine learning can also be applied to other domains, such as the security domain, since most of them are not data dependent attacks. Adversarial machine learning studies in the security domain (e.g., intrusion detection) received more and more attention in recent years. To understand the current thinking of adversarial machine learning in the security domain, we conduct a literature review of papers during the last decade. We use google scholar to trace publications and their citations and search papers that fall into this topic.

Table~\ref{tbl:securityStudy} lists relevant work from the last decade. We observe that researchers are more interested in attacks than defense, while at the same time, malware detection and intrusion detection receive more attention than other tasks (e.g., spam filtering), which also motivate our study.

\begin{table*}[!t]
\centering
\caption{A list of highly cited (i.e., those with at least ten cites per year since publication) studies of adversarial machine learning in the security-sensitive tasks. This list of papers was found from the Google scholar using the search query, e.g., ``(adversarial machine learning) and (malware)'' and ``(adversarial machine learning) and (intrusion detection)'' and ``(adversarial machine learning) and (malicious)''. We only use papers in the last ten years (2010-2020). The count of citations is retrieved from Google Scholar on Sept 9th, 2020.}
\begin{tabular}{c|c|c|c|c}
\hline
\rowcolor[HTML]{ECF4FF} 
\textbf{Ref} & \textbf{Year} & \multicolumn{1}{c|}{\cellcolor[HTML]{ECF4FF}\textbf{Citation}} & \textbf{Tasks} & \textbf{Attack/Defense} \\ \hline
\cite{biggio2010multiple} & 2010 & 175 & Spam Filtering & \RIGHTcircle \\
\cite{zhou2012adversarial} & 2012 & 87 & Spam Filtering & \LEFTcircle \\
\cite{smutz2012malicious} & 2012 & 207 & Malicious PDF  & \LEFTcircle \\
\cite{biggio2013data} & 2013 & 91 & Program Malware Detection & \LEFTcircle \\
\cite{biggio2013evasion} & 2013 & 471 & Malicious PDF Detection & \LEFTcircle \\
\cite{maiorca2013looking} & 2013 & 101 & Malicious PDF Detection & \LEFTcircle \\
\cite{laskov2014practical} & 2014 & 239 & Malicious PDF Detection & \CIRCLE \\
\cite{DBLP:conf/ccs/BiggioRAWCGR14} & 2014 & 80 & Spam and Malware Detection & \LEFTcircle \\ 
\cite{DBLP:conf/ccs/BiggioRAWCGR14} & 2014 & 95 & Program Malware Detection & \LEFTcircle \\
\cite{zhang2015adversarial} & 2015 & 126 & Spam Filtering and Malicious PDF Detection & \RIGHTcircle \\ 
\cite{DBLP:conf/icml/XiaoBBFER15} & 2015 & 211 & Malicious PDF Detection & \RIGHTcircle \\
\cite{DBLP:conf/icml/XiaoBBFER15} & 2015 & 160 & Malicious PDF Detection & \LEFTcircle \\ 
~\cite{DBLP:journals/corr/GrossePM0M16} & 2016 & 229 & Android Malware Detection & \LEFTcircle \\ 
\cite{tramer2016stealing} & 2016 & 623 & Spam Filtering & \LEFTcircle \\

\cite{xu2016automatically} & 2016 & 243 & Malicious PDF Detection & \LEFTcircle \\ 
~\cite{DBLP:journals/corr/GrosseMP0M17} & 2017 & 255 & Android Malware Detection & \LEFTcircle \\ 
\cite{DBLP:journals/corr/abs-1709-03423} & 2017 & 57 & Android Malware Detection & \CIRCLE \\
~\cite{hu2017generating} & 2017 & 172 & Program Malware Detection & \LEFTcircle \\ 
\cite{dang2017evading} & 2017 & 50 & Malicious PDF Detection & \LEFTcircle \\ 
\cite{chen2018automated} & 2018 & 77 & Android Malware Detection & \LEFTcircle \\
\cite{demontis2019adversarial} & 2019 & 20 & Android Malware Detection & \LEFTcircle \\ 
\hline
\end{tabular}
\label{tbl:securityStudy}
\end{table*}

\subsection{Adversarial Defense Strategy}

Researchers have proposed various solutions against adversarial machine learning attacks. One way to categorize defense strategies is based on different phases in the machine learning pipeline. Another way is to divide into two types, i.e., \textit{reactive defenses} and \textit{proactive defenses}. The former focuses on detecting adversarial examples separately from a trained classifier, while the latter focuses on building classifiers that are robust to adversarial examples. In the following section, we provide a summary of existing defense methods especially against adversarial evasion attacks.
 
Defense during the testing phases includes \textit{adversarial training}, \textit{gradient masking}, \textit{defensive distillation}, and \textit{ensemble learning}. The idea of \textit{adversarial training}~\cite{DBLP:conf/iclr/TramerKPGBM18,DBLP:conf/iclr/NaKM18,DBLP:conf/iclr/MadryMSTV18,DBLP:conf/iclr/FarniaZT19,DBLP:conf/nips/ShafahiNG0DSDTG19,yin2019adversarial} is to build a ``golden'' dataset that ideally contains a set of curated attacks and normal data that are representative of the target system. The data is then used when training the model. Intuitively, if the model sees adversarial examples during training, its performance during prediction will be improved for adversarial examples generated in the same way. However, the problem with adversarial training is that it suffers from an optimized attack or adaptive attack, since this method only defends the model against the same attacks used to craft the examples originally included during training. 

\textit{Gradient masking}~\cite{papernot2018sok} is a technique that hides the model gradients to reduce model's sensitivity to adversarial examples. However, later work~\cite{DBLP:conf/icml/AthalyeC018} shows that the gradient masking tactic does not work because of the transferability property of adversarial examples. The attackers can still build a substitute model and transfer the attacks.

\textit{Defensive distillation}~\cite{papernot2016distillation} tries to generate a new model whose gradients are much smaller than the original undefended model. If gradients are very small, some gradient-based attacks are no longer useful, as the attacker would need great distortions of the input data to achieve a sufficient change in the loss function. However, this method was quickly proved to be ineffective~\cite{DBLP:journals/corr/CarliniW16}. With a slight modification to a standard attack, attackers can still find adversarial examples on distilled networks.

\textit{Ensemble learning}~\cite{dietterich2002ensemble} is another widely used defense mechanism, in which multiple classifiers are combined together to improve classifier robustness. For example, Biggio et. al.~\cite{biggio2010multiple} investigated ways to build a multiple classifier system (MCS) that improved the robustness of linear classifiers. They argued that randomization-based MCS construction techniques, such as bagging and random subspace method (RSM), were effective in improving classification accuracy. Tramer et. al.~\cite{DBLP:conf/iclr/TramerKPGBM18} introduced a technique called ensemble adversarial training that augments training data with perturbations transferred from other models to increase robustness. Kariyappa et. al.~\cite{kariyappa2019improving} showed that an ensemble of models with misaligned loss gradients can provide an effective defense against transfer-based attacks. Besides, some other studies~\cite{DBLP:conf/iclr/TramerKPGBM18,smutz2016tree,kantchelian2016evasion} also show ways to respond to adversarial evasion attack via ensemble learning.

However, some other prior results are not supportive of the use of ensemble learning
for defense purposes~\cite{zhang2020decision,zhang2018gradient,he2017adversarial,DBLP:conf/iclr/TramerKPGBM18,DBLP:journals/corr/PapernotMG16}. For example, Zhang et al.~\cite{zhang2020decision} show that a discrete-valued tree ensemble classifier can be easily evaded by adversarial inputs manipulated based only on the model decision outputs. Zhang et al.~\cite{zhang2018gradient} investigate the evasion attacks in a more practical setting where attackers do not know the details of classifier, but instead they may acquire only a portion of the labeled data or a replacement dataset for learning the target decision boundary. They argue that ensemble classifiers are not necessarily more robust under a least effort attack based on gradient descent. He et al~\cite{he2017adversarial} demonstrate that an adaptive adversary can create adversarial examples successfully with low distortion, hence implies that ensemble of weak defenses is not sufficient to provide strong
defense against adversarial examples. Papernot et al.~\cite{DBLP:journals/corr/PapernotMG16} find the transferability property of adversarial examples also make ensemble classifier less effective.

As mentioned in the introduction section, we argue that there is an issue with prior research using ensemble methods for defense purpose. Specifically, the exploration space for their ensemble systems was too small, while our reading of Table~\ref{tbl:hyperparameters} indicates that there are trillions of options to configure a security detection model (the building block of the ensemble system). The rest of this paper proposes a novel method that takes better advantage of that large space of options.

\subsection{Hyperparameter Optimization}\label{sec:hyperparameter}

Given a space of hyperparameter options like Table~\ref{tbl:hyperparameters}, hyperparameter optimization can be represented as follows:

\begin{equation}
    x^{\star} = \operatorname*{argmax}_{x \in \chi} f(x)
\end{equation}
Here $f(x)$ represents the objective to be maximize, e.g. recall or accuracy. $x^{\star}$ is the set of hyperparameters that produce the best score, while $x$ can be any value from domain $\chi$.

Existing hyperparameter optimization algorithms can mainly fall into three categories:

\begin{itemize}
    \item Exhaustive search of hyperparameter space;
    \item Using evolutionary algorithm;
    \item Utilizing Bayesian optimization.
\end{itemize}

The first category includes \textit{manual search}, \textit{grid search} and \textit{random search}. The search space of each hyperparameter is discretized, and the total search space is discretized as the Cartesian products of them. When using manual search, we choose some model hyperparameters based on our own experience. We then train the model, evaluate its performance and start the process again. This loop is repeated until a satisfactory score is found. The grid search~\cite{bergstra2011algorithms} algorithm would traverse all the configurations and select the best one, which is computationally costly and can easily suffer from the ``\textit{curse of dimensionality}''. As a variation of grid search algorithm, random search algorithm~\cite{bergstra2012random} randomly samples the configurations to reach a predefined fixed sampling density. With purely random sampling, the selected hyperparameters give a non-uniform sampling density of the search space. 

As for the second category, the \textit{evolutionary algorithm} in hyperparameter optimization~\cite{young2015optimizing} takes its inspiration from the process of natural selection. This algorithm allows a selective exploration of the operation range using fitness function that determines which is going to be the next point to be sampled.

Both of them are good options for simple optimization problems due to easy implementation of algorithms. However, for complex objective functions, both methods are relatively inefficient because there is no guarantee that they can find an optimal solution except if the configuration space is thoroughly searched. This is a problem if the evaluations of the objective function are not cheap, i.e., the models take a significant amount of time to run with lots of computational resources, and they are uninformed of the information gained from previous evaluations. For example, they do not choose the next hyperparameter based on previous results, which leads to wasting a large amount of time evaluating ``bad'' hyperparameters instead of focusing on the most promising hyperparameters.

\begin{figure}[!t]
\centering
\includegraphics[width=12cm]{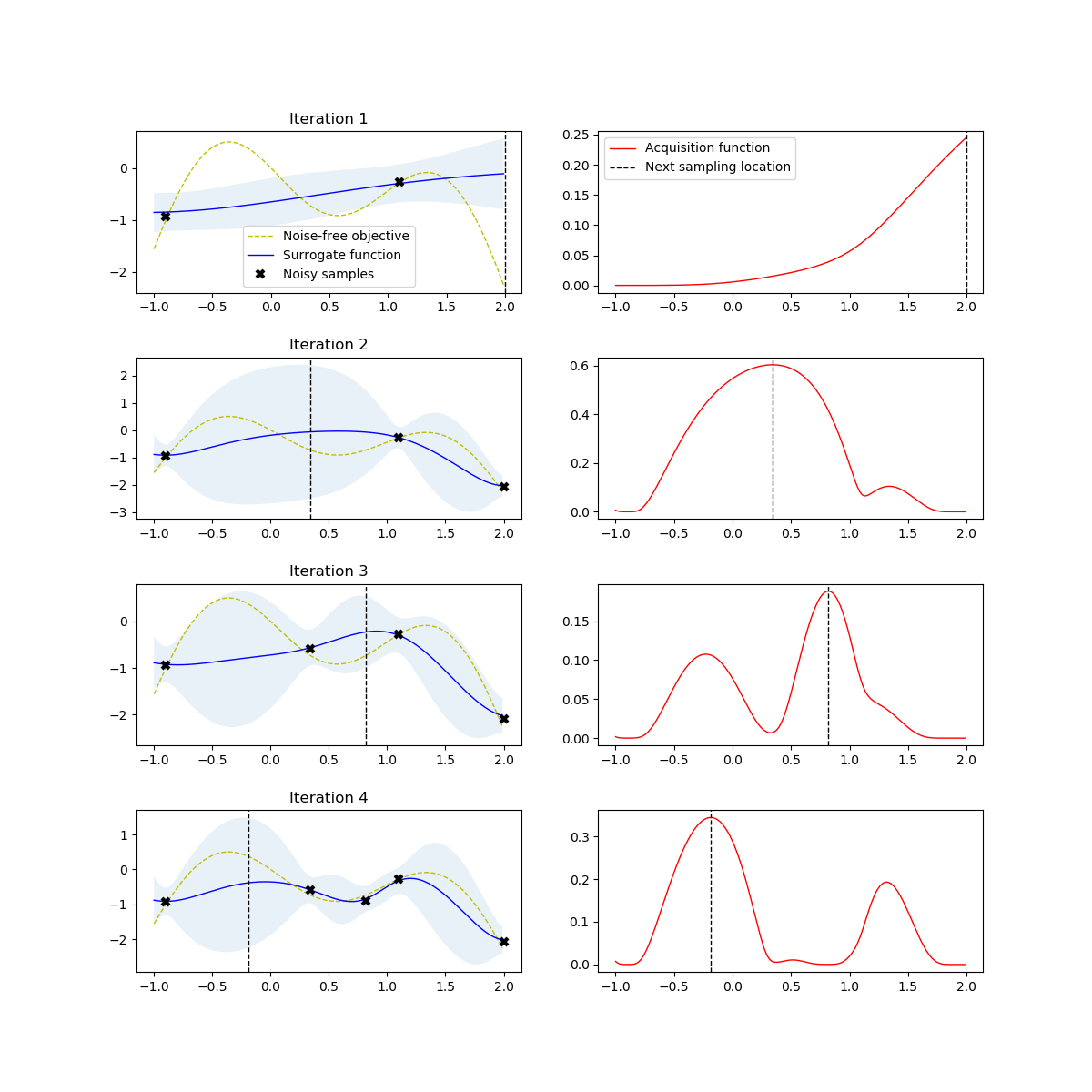}
\caption{An example of the Bayesian optimization process. Bayesian optimization incorporates prior belief about \textit{objective function} and updates the prior with samples drawn from objective function to get a posterior that better approximates objective function. The model used for approximating the objective function is called \textit{surrogate function}. Bayesian optimization also uses an \textit{acquisition function} that directs sampling to areas where an improvement over the current best observation is more likely.}
\label{fig:bayesian}
\end{figure}

\textit{Bayesian optimization}~\cite{snoek2012practical,shahriari2015taking} address such concerns by keeping track of past evaluation results. The principle of Bayesian optimization is using those results to build a probability model of objective function, and map hyperparameters to a probability of a score on the objective function, and therefore use it to select the most promising hyperparameters to evaluate in the true objective function. This method is also called \textit{Sequential Model-Based Optimization} (SMBO). Figure~\ref{fig:bayesian} shows an example process of Bayesian optimization.

The probability representation of the objective function is called \textit{surrogate function} or \textit{response surface} because it is a high-dimensional mapping of hyperparameters to the probability of a score on the objective function. The surrogate function is much easier to optimize than the objective function and Bayesian methods work by finding the next set of hyperparameters to evaluate the actual objective function by selecting hyperparameters that perform best on the surrogate function. This method continually updates the surrogate probability model after each evaluation of the objective function.

Variations of SMBO methods differ in how to build a surrogate of the objective function and the criterion used to select the next hyperparameters. Several choices for surrogates function are Gaussian Processes, Random Forest Regression, and Tree Parzen Estimators (TPE)~\cite{bergstra2011algorithms}, and one of the most common choices for acquisition function is Expected Improvement~\cite{snoek2012practical}. Specifically, our method chooses TPE as the surrogate function and Expected Improvement as the acquisition function, which is popular to use. TPE handles hyperparameter space in a tree-structured fashion, and during iterations, TPE divides observations into two groups. One group only contains observations that give the best scores after evaluation and the other group contains all the rest. The fraction of the best observation is usually defined as 10\% to 25\% of observations. TPE then models the likelihood probability of each group, and using the likelihood probability from the first group, TPE sampled a bunch of candidates, from which we seek a candidate that is more likely to be in the first group and less likely to be in the second group. TPE also uses the parzen-window density estimators, with which each sample defines Gaussian distribution with specified mean (i.e., the value of the hyperparameter) and standard deviation. These points then stack together and normalized to assure that output is Probability Density Function (PDF). For Expected Improvement, finding the values that will yield the greatest expected improvement in the surrogate function is much cheaper than evaluating the objective function itself.

\section{Methodology}~\label{sec:method}

In this section, we first introduce some guidelines that help direct our work. We then discuss the adversarial threat model. Next, we introduce the methods of generating adversarial examples for evasion attack. We then demonstrate the details of our ensemble learning based approach \textit{\textbf{Omni}}.

\subsection{Some Guidelines}

Before we go deeper into our proposed method, we first introduce several general design principles that are not specific to any individual research question. Carlini et al.~\cite{carlini2019evaluating} provide practical advices for evaluating adversarial defense that is intended to be robust to adversarial examples. Specifically, their work shows a list of principles of rigorous evaluations and a checklist of recommendations. Motivated by their work, here we list some example principles that are applicable to our work. Besides, we also add some other principles that beyond Carlini et al.'s recommendations.


\begin{itemize}
\item \textit{P1}: State a precise threat model that defense is supposed to be effective under.
\begin{itemize}
\item Our threat model is described in Section~\ref{sec:threat}.
\end{itemize}
\item \textit{P2}: Release models and source code.
\item \textit{P3}: Report clean model accuracy when not under attack.
\begin{itemize}
\item In the results section, we take care to present our pre-attack accuracy as well as other results.
\end{itemize}
\item \textit{P4}: Describe the attacks applied, and report all attack hyperparameters.
\item \textit{P5}: Apply a diverse set of attacks.
\item \textit{P6}: Report per-case attack success rate.
\item \textit{P7}: Record the runtime of hyperparameter tuning phase.
\item \textit{P8}: Any work with a stochastic component need to repeat experiments multiple times. As shown below, all our results come from multiple 80\% train, 20\% test runs where, for each run, the random number seed was changed.
\end{itemize}

\subsection{Threat Model}\label{sec:threat}

The machine learning adversarial threat model is a structured framework that lays out all the possible threat vectors on the machine learning system. We provide here a detailed threat model for evasion attacks against deep neural networks. This threat model consists of defining the adversary's goals, knowledge of the target system, and capabilities of manipulating the testing data.

\textit{\textbf{Adversary's goals.}} The goal of the adversary attackers is to impact target model's prediction performance by causing its misclassification in the testing phase, thus malicious payload can avoid being detected. For example, in a security system (e.g., intrusion detection or malware detection), the attackers want a specific class (e.g., ``malicious'') to be classified as a specific other class (e.g., ``benign'').

\textit{\textbf{Adversary's knowledge.}} The smart adversary attackers are essentially assumed to obtain the internal knowledge of the target model through other approaches. For example, we assume that attackers are able to collect information, including but not limited to model architecture, number of layers, optimization algorithm used, gradients of loss function, testing data, and therefore could successfully carry out white-box attacks~\cite{DBLP:journals/corr/GoodfellowSS14} to the target model. Prior study~\cite{DBLP:conf/ccs/Carlini017} has shown that accessing model's gradients is one of the most efficient ways to fool the target models with crafted adversarial examples. When a defense system is applied, the attackers are also assumed to obtain its knowledge and thus challenge the defense system.

\textit{\textbf{Adversary's capability.}} In adversarial evasion attacks, the adversary attackers are assumed to be able to perturb the testing dataset. In addition, the adversary attackers are also able to acquire prediction results from the target model. We assume adversarial attacks are carried out at the inference time (i.e., testing time), which means the attackers are only able to perturb the immediate inputs, while not be able to manipulate the training dataset.

Furthermore, we make some justifications about the threat model. Firstly, we limit the scope of study that the attackers will try to evade a single model with crafted adversarial examples. Attacking multiple models with adversarial examples~\cite{kwon2018multi} is another interesting research direction which we would like to explore in future work. In such type of attack, attackers generate multi-targeted adversarial examples which can be found useful for them to make multiple models to recognize a single data (e.g., image) as different classes.

Secondly, there is a growing interest in different types of privacy-related attacks (e.g., model extraction attack~\cite{papernot2017practical}) which make the leakage of model information possible~\cite{rigaki2020survey}. For example, in an example from the previous study~\cite{juuti2019prada}, some business models are hosted in a secure cloud that allow user clients to query the models via cloud-based prediction APIs. These prediction APIs are suffered from being exploited with model extraction attacks. The target model can be used as an oracle for returning predictions for the samples that attackers submit. Such kind of attempts can further be iteratively executed for attackers to maximize the information extraction about model internals.

Thirdly, our threat model also does not assume adaptive adversarial attacks which are more novel and specifically designed to target a given defense mechanism. In addition, the white-box adversarial attacks in this threat model are non-targeted attacks. Specifically, the aim of non-targeted attacks is to cause samples to be classified incorrectly while targeted attacks would cause samples to be misclassified as specific target class.

\subsection{Adversarial Attack Strategy}~\label{sec:attackStrategy}

\textit{Adversarial examples}~\cite{DBLP:journals/corr/SzegedyZSBEGF13} or \textit{adversarial inputs} are examples that are intentionally crafted by attackers by making small perturbations to the input data to cause a machine learning model to produce an incorrect output. Machine learning models, including existing state-of-the-art models such as neural networks lack the ability to classify adversarial examples correctly. A very active research field relevant to this topic is how to craft adversarial examples to fool models. We choose five methods that are widely studied in this domain.

\textit{\textbf{Fast Gradient Sign Method (FGSM)}}~\cite{DBLP:journals/corr/GoodfellowSS14} is a method to generate adversarial examples using model's gradient information. Each data in the clean dataset $x$ is modified by adding or subtracting an almost imperceptible error of $\epsilon$. If the sign of the gradient is positive, $\epsilon$ will be added, and vice versa. Equation~\ref{eqa:fgsm} formalizes the way to generate adversarial examples by FGSM. In this expression, we note that $x_{adv}$ is the adversarial dataset, $x$ is the original input dataset, $y$ is the original output label, $\epsilon$ is the multiplier to ensure the small perturbation, $\theta$ is the model hyperparameters, $\nabla$ is the gradient, and $\mathcal{L}$ is the loss function.

\begin{equation}\label{eqa:fgsm}
    x_{adv} = x + \epsilon* sign(\nabla_{x}\mathcal{L}(\theta, x, y))
\end{equation}
Under this assumption, the gradient of the loss function indicates the direction in which we need to change the input vector to produce a maximal change in the loss. In order to keep the size of the perturbation small, we only extract the sign of the gradient, not its actual norm, and scale it by a small factor epsilon.

\textit{\textbf{Basic Iterative Method (BIM)}}~\cite{KurakinGB17a} is an iterative version of FGSM where a small perturbation is added in each iteration. There are two versions of BIM attack: BIM-A and BIM-B. The BIM-A method stops the iteration as soon as mislcassification is achieved. In the BIM-B method, iterations only stop after a fixed number of rounds.

\textit{\textbf{Jacobian-based Saliency Map (JSMA)}}~\cite{papernot2016limitations} is an iterative method that achieves misclassification of input to any pre-specified class. It uses feature selection with the aim of minimizing the number of features perturbed (i.e., the $L_{0}$ distance metric).

This method includes the computation of saliency maps for an input sample, which contain the saliency values for each input feature. These values indicate how much the modification of that feature will perturb the classification process, how much different from each target class. Features are then selected in decreasing order of saliency value, and the feature with the max value in this map is perturbed by $\epsilon$. The saliency map is created in the following way:

\begin{equation}\label{eqa:jsma}
    S^{+}(x_{(i)}, c) = \begin{cases}
                        0 \quad  \text{if $\frac{\partial f(x)_{(c)}}{\partial x_{(i)}} < 0$ or ${\mathlarger\sum}_{c^{'}\neq c} \frac{\partial f(x)_{(c^{'})}}{\partial x_{(i)}} > 0$} \\
                        \\
                        -\frac{\partial f(x)_{(c)}}{\partial x_{(i)}} \cdot {\mathlarger\sum}_{c^{'}\neq c} \frac{\partial f(x)_{(c^{'})}}{\partial x_{(i)}} \quad \text{otherwise} \\
                        \end{cases}       
\end{equation}
We modify JSMA, so as to flip the feature values from 0 to 1, or 1 to 0 in the perturb step.

\textit{\textbf{DeepFool}}~\cite{moosavi2016deepfool} works in an iterative manner, with the aim of minimizing the euclidean distance between perturbed samples and original samples (i.e., the $L_{2}$ distance metric). Specifically, the generation of adversarial samples consists of the analytical calculation of a linear approximation of the decision boundary that separates sample from different classes, and then adding a perturbation perpendicular to that decision boundary, which brings the input closest to a linear approximation of the decision boundary. The algorithm terminates once misclassification is achieved.

\textit{\textbf{Carlini-Wagner (C \& W)}} attack~\cite{carlini2017towards,DBLP:conf/ccs/Carlini017} builds an optimization-based attack where the goal is to find the smallest perturbation that can cause a misclassification. If we consider input $x  \in [0,1]^{n}$ and noise $\delta \in [0,1]^{n}$, this attack finds the adversarial instance by finding the smallest noise $\delta \in \mathbb{R}^{n}$ added to the input $x$ that will change the classification to a class $t$. The noise level is measured in terms of $L_{p}$ distance. Finding the minimum $L_{p}$ distance of $\delta$ while ensuring that the output will have a class $t$ is not a trivial optimization problem. Instead, Carlini-Wagner attack solves the optimization problem of the form:

\begin{equation}
    \underset{\delta \in \mathbb{R}^{n}}{min} \quad ||\delta||_{p} + c \cdot f(x + \delta)
\end{equation}
where $x + \delta \in [0,1]^{n}$; $f$ is an objective function that drives the input $x$ to be misclassified; and  $c > 0$ is a suitably chosen constant. 

\subsection{\textit{\textbf{Omni}}: Building Ensembles of ``Unexpected'' Models}\label{sec:ensemble}

\subsubsection{System Architecture}\label{sec:architecture}

\begin{figure}[!ht]
\centering
\includegraphics[width=12cm]{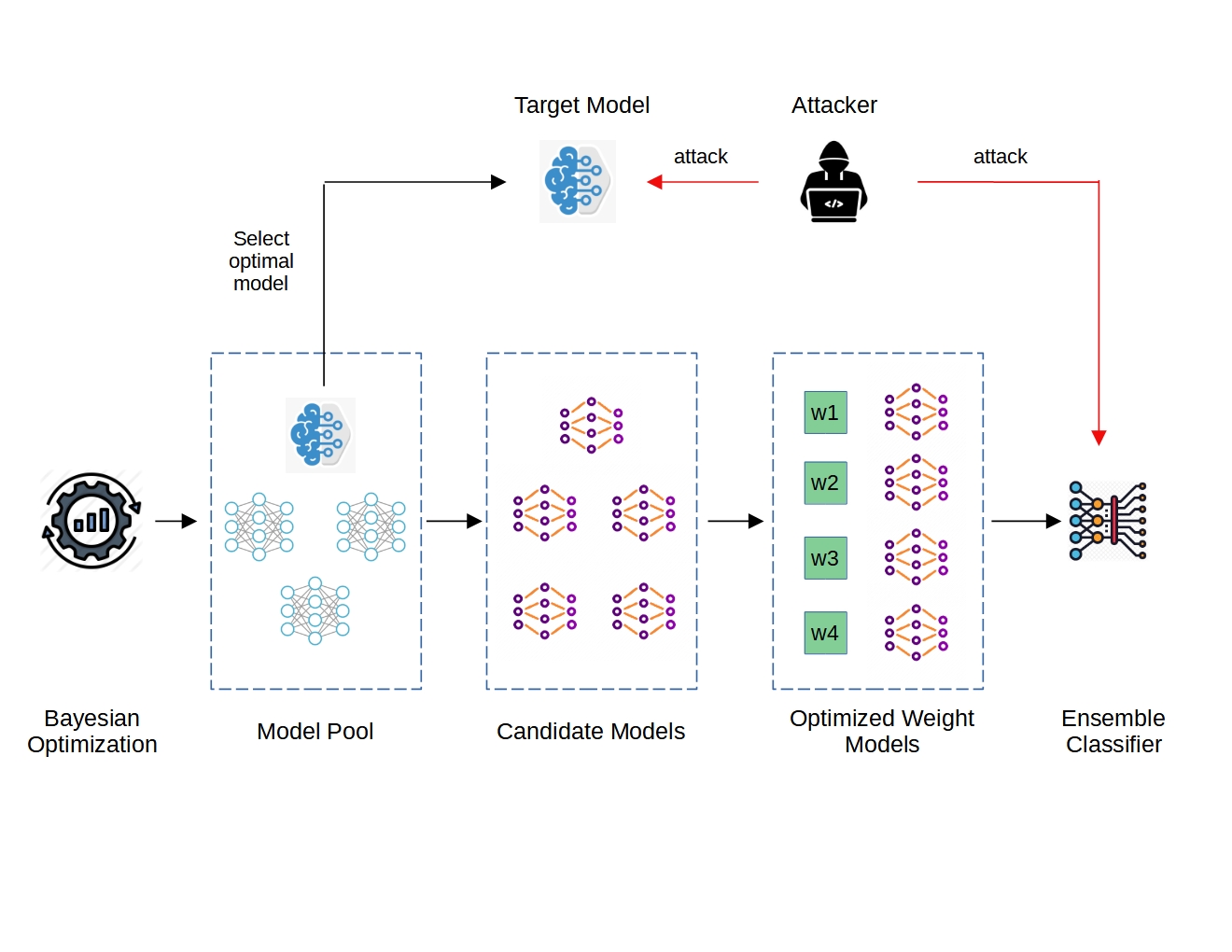}
\caption{The architecture of our proposed ensemble system.}
\label{fig:architecture}
\end{figure}

The basic idea of \textbf{\textit{Omni}} is to build a more robust ensemble classifier against adversarial attacks that aim at a target model. In order to create such robust ensemble classifier, we employ several novel tricks. The architecture of \textbf{\textit{Omni}} is presented in Figure~\ref{fig:architecture}. In our method, we first apply a hyperparameter tuning method named Bayesian optimization to train a set of models that build a model pool. We ranked the models in the model pool by prediction performance, and pick the model with optimal performance for normal prediction. This model is thus becoming the target model of the adversarial attackers.

As the defense strategy against the adversarial attacks, the ensemble classifier (i.e., \textbf{\textit{Omni}}) selects other candidate models from the model pool rather than the target model. Those candidate models are able to achieve sub-optimal prediction performance as well as with a distance (in terms of hyperparameter) to the target model. Those picked models are further optimized with weights in the ensemble classifier in order to perform better (i.e., robust) against adversarial attacks.

Besides, we assume that users are more likely to train their models and further optimize their models towards the goal of better prediction performance. We start with such point and choose the optimal model as the attacker's target since a sub-optimal model might suffer from misclassification even under normal prediction.

\subsubsection{Build Ensemble Classifier}~\label{sec:buildEnsemble}

Since {\bf Omni} is an ensemble learning based approach, we then come up with two questions: 1) \textit{How to select  models for the ensemble system?} and 2) \textit{How to aggregate  those models across the ensemble system?}

\textit{\textbf{Unexpected model selection.}} 
In our method,
we first run our optimizer in the usual manner to find parameters resulting in the model with the highest accuracy. We call this model the {\em expected model} since  
this is the model we would  expect the attackers to learn. 

Note that, as a side effect of this process, we have a large {\em pool} of models; i.e. all the other models explored by the optimizer before arriving at the best one. In this pool, we then hunt for ``useful'' {\em unexpected models}; i.e. models whose hyperparameters are {\em distant}
from the expected models. Here, by ``useful'' we mean the models that are performing nearly as well as the expected model. Using ensemble learning, we then create a combined classifier from the unexpected models. 

Why do we use those``unexpected models''?
Previous work~\cite{papernot2016transferability,demontis2019adversarial} show that adversarial evasion attack has the ability to transfer from one model to another model. That is, specially crafted adversarial examples that cause misprediction of model $A$ are also likely to mislead a different model $B$. However, this transferability becomes less reliable
when  two models have different structures~\cite{papernot2016transferability}. For example:
\bi
\item
For cross-technique models (e.g., SVM and DNN), attacks from SVM to DNN are less effective than from SVM to SVM. 
\item
For intra-technique models (e.g., both models are DNN), attacks from DNN $A$ to DNN $B$ are also less effective than from DNN $A$ to itself.
\item
Generalizing these examples, the rest of
this paper tests that we can generate different structures (that makes an adversary harder to figure out defender's decision boundary), just be picking models that are some distance ``$t$''
from the expected model (measured using the distance between the hyperparameters.
\ei

To implement the above, we need some measure of ``distance'' $t$, that denotes an ``unexpected model''.
To compute that, we note that   the hyperparameters in Table~\ref{tbl:hyperparameters} are in various forms, which involve both discrete categorical and continuous numerical values. Traditional metrics such as Euclidean distance or cosine distance cannot calculate distance between two entities whose attributes have a mix of values. \textit{Gower distance}~\cite{gower1971general} is a distance measure that can be used to address this concern. This measure is defined as follows:

{\small \begin{equation}
    d(i,j) = \frac{\sum_{k=1}^{N} w_{ij}^{(k)}d_{ij}^{(k)}}{\sum_{k=1}^{N} w_{ij}^{(k)}}
\end{equation}}
where $w_{ij}^{(k)}$ is the weight of variable $k$ between observation $i$ and $j$ and $d_{ij}^{(k)}$ is the distance between $i$ and $j$ on variable $k$. Moreover, $d_{ij}^{(k)}$ applies different formulas to categorical and numerical variables. Specifically, for categorical variable, if two observations are the same then the distance $d_{ij}^{(k)}$ of them is assigned $0$, otherwise assigned $1$. For numerical variable, the absolute difference is calculated between observation $i$ and $j$, then the result is scaled to range $[0,1]$ by dividing the range of values of variable $k$. The following equations describe how to calculate categorical and numerical variables respectively, where $x_{i}^{(k)}$ is the value of variable $k$ for observation $i$.

{\small \begin{equation}
    d_{ij}^{(k)} =  
    \begin{cases}
    1, & \text{if} \: x_{i}^{(k)} \neq x_{j}^{(k)}\\
    0, & \text{if} \: x_{i}^{(k)} = x_{j}^{(k)}
    \end{cases}
\end{equation}  

\begin{equation}
    d_{ij}^{(k)} = \frac{|x_{i}^{(k)} - x_{j}^{(k)}|}{max(x^{(k)}) - min(x^{(k)})}
\end{equation}}

 The Gower distance allows to assign a weight $w_{ijk}$ to each individual variable base on the importance of that variable in the distance calculation. For simplicity, we use the equal weight in our study.


\textit{\textbf{Unexpected model aggregation.}} There are various ways of combining unexpected models, in our work, \textit{\textbf{Omni}} uses a bagging approach named \textit{weighted ensemble}. When optimizing these weights,
we seek to maximize the accuracy of the ensemble.

\begin{figure}[!htbp]
\centering
\scriptsize 
\begin{center}
\begin{minipage}{2.5in}\begin{lstlisting}[mathescape,linewidth=7.5cm,frame=none,numbers=none]
  def DE(np=20, cf=0.75, f=0.3, lives=10):  # default settings
    frontier = # make "np" number of random guesses
    best = frontier.1 # any value at all
    while(lives$--$ > 0): 
      tmp = empty
      for i = 1 to $|$frontier$|$: # size of frontier
         old = frontier$_i$
         x,y,z = any three from frontier, picked at random
         new= copy(old)
         for j = 1 to $|$new$|$: # for all attributes
           if rand() < cf    # at probability cf...
              new.j = $x.j + f(z.j - y.j)$  # ...change item j
         # end for
         new  = new if better(new,old) else old
         tmp$_i$ = new 
         if better(new,best) then
            best = new
            lives++ # enable one more generation
         end
      # end for
     frontier = tmp
     lives--
    # end while
    return best
\end{lstlisting} 
\end{minipage}
\end{center}
\caption{Pseudocode of differential evolution. Extended from~\cite{storn1997differential}.}
\label{fig:pseudo_DE} 
\end{figure}

The way of searching for the weight values is an optimization process. We propose to use an evolutionary algorithm named \textit{differential evolution} (DE), shown in Figure~\ref{fig:pseudo_DE},  to solve the function optimization. This optimizer is inspired by biology natural selection process and follows the rule of Darwin's ``\textit{survival of the fitness}'' evolution theory. The general idea of this algorithm is to randomly select some samples from a population. The selected samples are then combined with a scheme to generate a new sample. If the new sample is better after evaluation with target function, then the previous samples are replaced. After several iterations, the population will converge towards the optimal solution. Differential evolution algorithm has very few parameters to adjust, so it is easy to implement and convenient to use.

The premise of the Figure~\ref{fig:pseudo_DE} code is that the best way to mutate the existing optimizations is to extrapolate between current solutions (stored in the {\em frontier} list). Three solutions $x, y, z$ are selected at random from the {\em frontier}. For each tuning parameter $j$, at some probability $cf$, DE replaces the old solution $x_j$ with {\em new}  where \mbox{$\mathit{new}_j = x_j + f \times (y_j - z_j)$} and $f$ is a parameter controlling differential weight.  

The main loop of DE runs over the {\em frontier} of size $np$, replacing old population with new candidates (if new candidate is better). This means, as the loop progresses, the {\em frontier}  contains increasingly more valuable solutions (which in turn helps extrapolation since the next time we pick $x,y,z$, we get better candidates). DE's loops keep repeating until it runs out of {\em lives}. The number of {\em lives} is decremented for each loop (and  incremented every time we find a better solution).

Note that this paper has {\em two} optimization problems requiring different optimization technologies.  Exploring the hyperparameters is a complex task which, if done naively, results in very slow optimization runtime. For that reason, as described above, we optimize the values of Table~\ref{tbl:hyperparameters} very carefully using the TPE-based Bayesian optimization method~\cite{bergstra2011algorithms} 
On the other hand, once a deep learner is tuned,
then its runtime for making predictions is fast. Hence, for just fiddling with the weights placed on the conclusions of a deep learner, we use the much simpler optimizer of  Figure~\ref{fig:pseudo_DE},

\section{Experiments}~\label{sec:experiment}
\subsection{Datasets}

To assess {\bf Omni}, we use the five security datasets of Table~\ref{tbl:dataOverview}, which covers various attack types including network traffic, Android malwares and malicious PDF files. These datasets were selected since they are publicly available and widely used in the security literature.  

\begin{table}[!h]
\centering
\small
\caption{An overview of the statistics of the security datasets studied in our work.}
\begin{tabular}{l|r|c|c}
\hline
\rowcolor[HTML]{ECF4FF} 
\multicolumn{1}{c|}{\cellcolor[HTML]{ECF4FF}\textbf{Dataset}} & \multicolumn{1}{c|}{\cellcolor[HTML]{ECF4FF}\textbf{Original Size}} & \textbf{Sampling Rate(\%)} & \textbf{Feature Count} \\ \hline
NSL-KDD & 148,517 & 100 & 123 \\ 
CSE-CIC-IDS2018 & 16,233,003 & 5 & 70 \\ 
CIC-IDS-2017 & 2,830,743 & 20 & 70 \\ 
CICAndMal2017 & 2,618,533 & 20 & 71 \\ 
Contagio PDF Malware & 22,525 & 100 & 135 \\ \hline
\end{tabular}
\label{tbl:dataOverview}
\end{table}

\textit{NSL-KDD}~\cite{nsl-kdd} dataset is an improved version of KDD'99 dataset~\cite{tavallaee2009detailed}, which recorded network traffic under different types of attacks. Compared with the original KDD dataset, NSL-KDD dataset removes redundant records in the train set and test set, which reduces the bias of trained classifiers towards the frequent records and further improves the detection rates. 

\textit{CIC-IDS-2017}~\cite{sharafaldin2018toward} is a dataset that consists of labeled network flows. It is comprised of both normal traffic and simulated abnormal data caused by intentional attacks on a test network. This dataset was constructed using the NetFlowMeter Network Traffic Flow analyzer, which collected more than 80 network traffic features and supported Bi-directional flows.

\textit{CSE-CIC-IDS2018}~\cite{sharafaldin2018toward} is another intrusion detection dataset collected in 2018 by Canadian Institute for Cybersecurity (CIC) on AWS (Amazon Web Services). This dataset includes seven different attack scenarios such as Brute-force, Heartbleed, Botnet, DoS, DDoS, Web attacks, and infiltration of the network from inside. Using the tool CICFlowMeter-V3, this dataset includes the captured network traffic and system logs of each machine.

\textit{CICAndMal2017}~\cite{lashkari2018toward} is an Android malware dataset that collects 426 malicious and 1,700 benign applications collected from 2015 to 2017 by researchers at the University of New Brunswick (UNB). The malicious samples are split into four categories (Adware, Ransomware, Scareware, SMS Malware) and 42 families. In addition to providing the APK files, the authors also ran each malicious sample on real Android smartphones and captured network traffic during installation, before restart, and after restart.

\textit{Contagio PDF Malware}~\cite{contagio-pdf} dataset is widely available and used for signature research and testing. This source of datasets was selected because it contained a large number of labeled benign and malicious PDF documents, including a relatively large number from targeted attacks.

Note that some datasets (e.g., CSE-CIC-IDS2018) have millions of entries, which would significantly increase the computational cost of both model training and testing on deep neural network. To simplify our evaluation, we apply the stratified random sampling strategy with pre-defined sampling rate. In this way, we maintain the imbalanced characteristic of security datasets (see Table~\ref{tbl:dataInPhase}).

\begin{table}[!h]
\centering
\small
\caption{The characteristics of security datasets during training and testing phase.}
\begin{tabular}{l|r|r|r|r|r|r}
\hline
\rowcolor[HTML]{ECF4FF} 
\multicolumn{1}{c|}{\cellcolor[HTML]{ECF4FF}} & \multicolumn{2}{c|}{\cellcolor[HTML]{ECF4FF}\textbf{Training Phase}} & \multicolumn{2}{c|}{\cellcolor[HTML]{ECF4FF}\textbf{Testing Phase}} & \multicolumn{2}{c}{\cellcolor[HTML]{ECF4FF}\textbf{Total}} \\ \cline{2-7} 
\rowcolor[HTML]{ECF4FF} 
\multicolumn{1}{c|}{\multirow{-2}{*}{\cellcolor[HTML]{ECF4FF}\textbf{Dataset}}} & \multicolumn{1}{c|}{\cellcolor[HTML]{ECF4FF}\textbf{Benign}} & \multicolumn{1}{c|}{\cellcolor[HTML]{ECF4FF}\textbf{Malicious}} & \multicolumn{1}{c|}{\cellcolor[HTML]{ECF4FF}\textbf{Benign}} & \multicolumn{1}{c|}{\cellcolor[HTML]{ECF4FF}\textbf{Malicious}} & \multicolumn{1}{c|}{\cellcolor[HTML]{ECF4FF}\textbf{Benign}} & \multicolumn{1}{c}{\cellcolor[HTML]{ECF4FF}\textbf{Malicious}} \\ \hline
NSL-KDD & 67,343 & 58,530 & 12,833 & 9,711 & 80,176 & 68,341 \\ 
CSE-CIC-IDS2018 & 535,701 & 102,379 & 133,926 & 25,595 & 669,627 & 127,974 \\ 
CIC-IDS-2017 & 363,410 & 89,050 & 90,853 & 22,262 & 454,263 & 111,312 \\ 
CICAndMal2017 & 193,777 & 224,873 & 48,445 & 56,218 & 242,222 & 281,091 \\ 
Contagio PDF Malware & 8,821 & 9,199 & 2,205 & 2,300 & 11,026 & 11,499 \\ \hline
\end{tabular}
\label{tbl:dataInPhase}
\end{table}

The sampled datasets are further pre-processed with the one-hot-encoding technique that encodes all categorical features to one-hot numeric array. We also apply the StandardScaler pre-processor to transform the data such that their distribution will have a mean value $0$ and standard deviation of $1$. Both of the procedures are implemented with the Scikit-learn toolkits. We also note that our task is essentially a binary classification problem, and we do not distinguish the attack types (e.g., Denial of Service) in the datasets. All data originally labeled with the type of attacks are labeled as malicious, and the others are labeled as benign.

\subsection{Experiment Rigs}

In all our experiments, 
we split the sampled dataset into two parts:
\bi
\item
Part 1: 80\% of the sampled dataset is used for   training and   optimization.
\item
Part 2: the rest of the sampled data is used for model testing.
\ei
The first part of the data ((Part 1)) is further split with ratio 3:1. During each trail of Bayesian optimization, the model is trained with the former part (75\%), and then evaluated in the latter part (25\%).

As mentioned above, for fast learners, these splits are typically created ten times. However,
for deep learning experiments (like this paper), since the training times are so long, three samples are often used.

We have referred to existing open-source CleverHans~\cite{papernot2016technical} library during the implementation of adversarial attacks. In addition, experiments are implemented on the Tensorflow framework, and conducted on our university's ARC cluster~\cite{ARC}, which provides Nvidia GPU computing resources. Experiments are also repeated multiple times (i.e., 20) (with different random number seeds), and median results are shown.

\subsection{Adversarial Attack Parameters}

Table~\ref{tbl:attackParameters} lists the parameters that we set for each type of adversarial attack. Our experiment results in Section~\ref{sec:result} are based on these default settings. In the table, the epsilon parameter indicates the degree of perturbation, and the clip function is defined as follows:

\begin{table}[!htbp]
\centering
\caption{The default attack parameters used in our experiment.}
\begin{tabular}{|c|l|}
\hline
\rowcolor[HTML]{ECF4FF} 
\textbf{Attack} & \multicolumn{1}{c|}{\cellcolor[HTML]{ECF4FF}\textbf{Parameters}} \\ \hline
FGSM & epsilon: 0.2, clip\_min: 0.0, clip\_max: 1.0 \\ \hline
\begin{tabular}[c]{@{}c@{}}BIM-A\\ BIM-B\end{tabular} & \begin{tabular}[c]{@{}l@{}}epsilon: 0.2, clip\_min: 0.0, clip\_max: 1.0, \\ iterations: 10\end{tabular} \\ \hline
DeepFool & epsilon: 0.2, clip\_min: 0.0, clip\_max: 1.0 \\ \hline
JSMA & \begin{tabular}[c]{@{}l@{}}theta: 1.0, gamma: 1.0, clip\_min: 0.0, \\ clip\_max: 1.0\end{tabular} \\ \hline
Carlini-Wagner & Iteration: 100, clip\_min: 0.0, clip\_max: 1.0 \\ \hline
\end{tabular}
\label{tbl:attackParameters}
\end{table}

\begin{equation}
  clip(x) =
    \begin{cases}
      MIN & \text{if $x < MIN$}\\
      x & \text{if $ MIN \leq x \leq MAX$}\\
      MAX & \text{if $x > MAX$}
    \end{cases}       
\end{equation}

~\newpage
\section{Results}~\label{sec:result}

\subsection{Evaluation Treatments}~\label{sec:treatment}

In our experiment, we evaluate our proposed approach with several treatments below:

\begin{itemize}
    \item \textit{Treatment 0 : Normal prediction}. In this treatment, we do not apply any adversarial attack, therefore models are trained on training datasets and normal predictions are made on testing datasets.
    \item \textit{Treatment 1 : Adversarial attacks}. This treatment creates adversarial examples with strategies introduced in Section~\ref{sec:attackStrategy}, with which the testing datasets are perturbed. Trained models then make prediction on the modified testing datasets.
    \item \textit{Treatment 2 : Adversarial training}. Previous works~\cite{DBLP:conf/iclr/TramerKPGBM18,DBLP:conf/iclr/NaKM18,DBLP:conf/iclr/MadryMSTV18,DBLP:conf/iclr/FarniaZT19} proposed a simple and intuitive strategy to train an adversarially robust model called ``adversarial training'', the basic idea of which is to produce adversarial examples and then incorporate adversarial examples into the training process. The robustness achieved by adversarial training depends on the strength of the adversarial examples used. 
    In this treatment, for each type of adversarial attack, we generate adversarial examples with attack parameter set $S_1$ on a portion of testing datasets, then we aggregate training datasets with created adversarial examples. We then perform same type of adversarial attack with adversarial examples generated with attack parameter set $S_2$ while $S_1 \neq S_2$. With ``adversarial training'', the trained models are expected to learn some traits of existing adversarial examples in order to make better predictions on new adversarial examples.
    \item \textit{Treatment 3 : Attack random ensemble}. In this treatment, we build an ensemble classifier that does not apply any specific strategy of model selection but random pick within a pool of models generated from Bayesian optimization. This random strategy neither guarantees the quality of models (in terms of performance), nor applies the distance criteria. Adversarial attacks are applied to the ensemble classifier.
    \item \textit{Treatment 4 : Attack average weight ensemble}. In this treatment, we build an ensemble classifier by selecting ``useful'' \textit{unexpected models} with methods introduced in Section~\ref{sec:buildEnsemble}. Models in the ensemble system are with a scheme that all weights of constitute models are equal, i.e., the final prediction is the average across the ensemble system. Adversarial attacks are applied on the ensemble classifier.
    \item \textit{Treatment 5 : Attack \textbf{\textit{Omni}}}. Same as \textit{Treatment 4}, \textbf{\textit{Omni}} selects models with defined strategies. Rather than enforcing average weights in the ensemble system, the weights of models are optimized with the ``differential evolution'' algorithm that tries to maximize the performance of the ensemble classifier under adversarial attacks.
\end{itemize}

\subsection{Treatment Evaluation Results}~\label{sec:treatmentResults}

We present the empirical experiment results of different treatments from Table~\ref{tbl:contagioAccuracy} to Table~\ref{tbl:CICAndMal2017Accuracy}. Each table is a summary result of an individual dataset. We also denote $A_0, A_1, A_2, A_3, A_4$ and $A_5$ as the accuracy of each treatment, respectively. Note that the $A_0$ results are shown in the header of each result table. We now discuss the observations from these tables.

Accuracy results $A_0$ in \textit{Treatment 0} indicate that prior to the adversarial evasion attacks, the step of hyperparameter optimization lets us find models with very high prediction performance of about 85\% to 99\% across all datasets. Then sophisticated adversarial attacks are proved to be effective in dropping those accuracy by a large amount (as shown in $A_1$ of \textit{Treatment 1}).

\begin{table}[!b]
\centering
\caption{Classification \textit{accuracy} (\%) on adversarial examples of dataset Contagio PDF Malware. The normal accuracy of the dataset of \textit{Treatment 0} ($A_0$) is \textit{99.64}\%. For \textbf{\textit{Omni}}, the distance \textit{d} = 0.9 is used.}
\begin{tabular}{|c|c|c|c|c|c|c}
\hline
\textbf{Attacks} &
  \textbf{\begin{tabular}[c]{@{}c@{}}\textit{Treatment 1} \\ ($A_1$)\end{tabular}} &
  \textbf{\begin{tabular}[c]{@{}c@{}}\textit{Treatment 2} \\ ($A_2$)\end{tabular}} &
  \textbf{\begin{tabular}[c]{@{}c@{}}\textit{Treatment 3}\\ ($A_3$)\end{tabular}} &
  \textbf{\begin{tabular}[c]{@{}c@{}}\textit{Treatment 4}\\ ($A_4$)\end{tabular}} & 
  \textbf{\begin{tabular}[c]{@{}c@{}}\textit{Treatment 5}\\ ($A_5$)\end{tabular}} \\ \hline
\hline
FGSM     & 37.58 & 59.33 & 33.67 & 57.53 & \textbf{70.48} \\ \hline
BIM-A    & 14.24 & 55.25 & 22.95 & 51.18 & \textbf{58.72} \\ \hline
BIM-B    & 35.67 & 60.71 & 36.53 & 52.54 & \textbf{65.03} \\ \hline
JSMA     & 62.65 & 72.86 & 56.82 & 70.36 & \textbf{76.22} \\ \hline
DeepFool & 67.46 & 71.23 & 55.42 & 73.42 & \textbf{87.37} \\ \hline
C\&W     & 55.42 & 66.49 & 45.92 & 63.55 & \textbf{74.41} \\ \hline
\end{tabular}
\label{tbl:contagioAccuracy}
\end{table}

\begin{table}[!htbp]
\centering
\caption{Classification \textit{accuracy} (\%) on adversarial examples of dataset NSL-KDD. The normal accuracy of the dataset of \textit{Treatment 0} ($A_0$) is \textit{84.82}\%. For \textbf{\textit{Omni}}, the distance \textit{d} = 0.9 is used.}
\begin{tabular}{|c|c|c|c|c|c|c}
\hline
\textbf{Attacks} &
  \textbf{\begin{tabular}[c]{@{}c@{}}\textit{Treatment 1} \\ ($A_1$)\end{tabular}} &
  \textbf{\begin{tabular}[c]{@{}c@{}}\textit{Treatment 2} \\ ($A_2$)\end{tabular}} &
  \textbf{\begin{tabular}[c]{@{}c@{}}\textit{Treatment 3}\\ ($A_3$)\end{tabular}} &
  \textbf{\begin{tabular}[c]{@{}c@{}}\textit{Treatment 4}\\ ($A_4$)\end{tabular}} & 
  \textbf{\begin{tabular}[c]{@{}c@{}}\textit{Treatment 5}\\ ($A_5$)\end{tabular}} \\ \hline
\hline
FGSM     & 56.87 & 70.54 & 48.12 & 69.12 & \textbf{79.14} \\ \hline
BIM-A    & 55.64 & 72.17 & 56.67 & 67.10 & \textbf{74.52} \\ \hline
BIM-B    & 63.87 & 69.46 & 61.72 & 66.91 & \textbf{74.65} \\ \hline
JSMA     & 47.32 & 54.92 & 48.62 & 53.82 & \textbf{73.17} \\ \hline
DeepFool & 57.03 & 71.21 & 56.13 & 76.74 & \textbf{83.22} \\ \hline
C\&W     & 44.87 & 67.26 & 47.21 & 66.79 & \textbf{71.14} \\ \hline
\end{tabular}
\label{tbl:nslAccuracy}
\end{table}

\begin{table}[!htbp]
\centering
\caption{Classification \textit{accuracy} (\%) on adversarial examples of dataset CIC-IDS-2017. The normal accuracy of the dataset of \textit{Treatment 0} ($A_0$) is \textit{92.56}\%. For \textbf{\textit{Omni}}, the distance \textit{d} = 0.9 is used.}
\begin{tabular}{|c|c|c|c|c|c|c}
\hline
\textbf{Attacks} &
  \textbf{\begin{tabular}[c]{@{}c@{}}\textit{Treatment 1} \\ ($A_1$)\end{tabular}} &
  \textbf{\begin{tabular}[c]{@{}c@{}}\textit{Treatment 2} \\ ($A_2$)\end{tabular}} &
  \textbf{\begin{tabular}[c]{@{}c@{}}\textit{Treatment 3}\\ ($A_3$)\end{tabular}} &
  \textbf{\begin{tabular}[c]{@{}c@{}}\textit{Treatment 4}\\ ($A_4$)\end{tabular}} & 
  \textbf{\begin{tabular}[c]{@{}c@{}}\textit{Treatment 5}\\ ($A_5$)\end{tabular}} \\ \hline
\hline
FGSM     & 40.29 & 62.47 & 51.03 & 63.07 & \textbf{77.13} \\ \hline
BIM-A    & 57.12 & 77.61 & 63.18 & 64.16 & \textbf{73.57} \\ \hline
BIM-B    & 53.58 & 67.89 & 48.94 & 70.18 & \textbf{76.24} \\ \hline
JSMA     & 40.15 & 58.62 & 42.13 & 52.27 & \textbf{74.31} \\ \hline
DeepFool & 50.18 & 63.18 & 44.91 & 65.31 & \textbf{76.52} \\ \hline
C\&W     & 46.23 & 61.34 & 42.86 & 59.64 & \textbf{70.35} \\ \hline
\end{tabular}
\label{tbl:CICIDS2017Accuracy}
\end{table}

The $A_2$ results show the effectiveness of \textit{Treatment 2} (i.e., adversarial training). In our study, we see that this strategy is beneficial in building a more robust model (compared to the results of $A_1$). The pre-trained models learn some traits of adversarial perturbations, and therefore are able to correctly classify more testing data when facing new adversarial examples. Note that in our study, models are adversarially trained with adversarial examples within the same adversarial attack type (i.e., attack specific), and therefore such strategy might be less effective in face of a new class of adversarial attacks or optimized attacks that would generate stronger perturbations. To address these concerns, several other studies proposed different adversarial training schemes such as ``ensemble adversarial training''~\cite{DBLP:conf/iclr/TramerKPGBM18} which augments training data with perturbations transferred from other models. We argue that these methods can be further explored as one of our future directions.

As to the $A_3$ results from \textit{Treatment 3}, these results show no fixed patterns for adversarial defense. We can see that without specific strategy of model selection, in some cases such as FGSM attack on the NSL-KDD dataset, the prediction accuracy are close to or even slightly lower than $A_1$ results. Several factors may contribute to these mixed results, such as using weak models to constitute the ensemble system or using a similar model to the victim model that enables transferability attacks, or both. These results could further suggest the desire to build a well-designed ensemble system rather than a random schema.

\begin{table}[!htbp]
\centering
\caption{Classification \textit{accuracy} (\%) on adversarial examples of dataset CSE-CIC-IDS2018. The normal accuracy of the dataset of \textit{Treatment 0} ($A_0$) is \textit{94.48}\%. For \textbf{\textit{Omni}}, the distance \textit{d} = 0.9 is used.}
\begin{tabular}{|c|c|c|c|c|c|c}
\hline
\textbf{Attacks} &
  \textbf{\begin{tabular}[c]{@{}c@{}}\textit{Treatment 1} \\ ($A_1$)\end{tabular}} &
  \textbf{\begin{tabular}[c]{@{}c@{}}\textit{Treatment 2} \\ ($A_2$)\end{tabular}} &
  \textbf{\begin{tabular}[c]{@{}c@{}}\textit{Treatment 3}\\ ($A_3$)\end{tabular}} &
  \textbf{\begin{tabular}[c]{@{}c@{}}\textit{Treatment 4}\\ ($A_4$)\end{tabular}} & 
  \textbf{\begin{tabular}[c]{@{}c@{}}\textit{Treatment 5}\\ ($A_5$)\end{tabular}} \\ \hline
\hline
FGSM     & 61.59 & 68.89 & 55.68 & 74.31 & \textbf{86.68} \\ \hline
BIM-A    & 49.61 & 77.67 & 56.79 & 70.54 & \textbf{85.03} \\ \hline
BIM-B    & 66.33 & 82.17 & 70.02 & 74.15 & \textbf{84.53} \\ \hline
JSMA     & 56.09 & 63.89 & 51.17 & 63.41 & \textbf{77.87} \\ \hline
DeepFool & 66.12 & 64.73 & 52.19 & 70.38 & \textbf{87.12} \\ \hline
C\&W     & 54.07 & 63.39 & 51.05 & 61.73 & \textbf{75.23} \\ \hline
\end{tabular}
\label{tbl:CICIDS2018Accuracy}
\end{table}

\begin{table}[!htbp]
\centering
\caption{Classification \textit{accuracy} (\%) on adversarial examples of dataset CICAndMal2017. The normal accuracy of the dataset of \textit{Treatment 0} ($A_0$) is \textit{95.47}\%. For \textbf{\textit{Omni}}, the distance \textit{d} = 0.9 is used.}
\begin{tabular}{|c|c|c|c|c|c|c}
\hline
\textbf{Attacks} &
  \textbf{\begin{tabular}[c]{@{}c@{}}\textit{Treatment 1} \\ ($A_1$)\end{tabular}} &
  \textbf{\begin{tabular}[c]{@{}c@{}}\textit{Treatment 2} \\ ($A_2$)\end{tabular}} &
  \textbf{\begin{tabular}[c]{@{}c@{}}\textit{Treatment 3}\\ ($A_3$)\end{tabular}} &
  \textbf{\begin{tabular}[c]{@{}c@{}}\textit{Treatment 4}\\ ($A_4$)\end{tabular}} & 
  \textbf{\begin{tabular}[c]{@{}c@{}}\textit{Treatment 5}\\ ($A_5$)\end{tabular}} \\ \hline
\hline
FGSM     & 44.41 & 56.33 & 42.38 & 63.91 & \textbf{74.93} \\ \hline
BIM-A    & 15.19 & 34.43 & 19.17 & 37.28 & \textbf{48.48} \\ \hline
BIM-B    & 43.09 & 57.28 & 46.39 & 54.11 & \textbf{71.72} \\ \hline
JSMA     & 45.67 & 56.07 & 46.91 & 53.38 & \textbf{72.35} \\ \hline
DeepFool & 55.32 & 57.76 & 46.82 & 62.53 & \textbf{76.21} \\ \hline
C\&W     & 43.54 & 59.43 & 44.72 & 57.26 & \textbf{68.47} \\ \hline
\end{tabular}
\label{tbl:CICAndMal2017Accuracy}
\end{table}

$A_4$ results of \textit{Treatment 4} come from a simple ensemble weighting scheme where all weights of constitute models are equal (so the final conclusion is the average across the ensemble). We note that our model selection strategy shows benefits with building a better robust model (compared with \textit{Treatment 3}). Furthermore, we note that,  this approach always performs worse than our proposed \textit{\textbf{Omni}}'s $A_5$ results in \textit{Treatment 5}. This is due to the use of differential evolution in \textit{\textbf{Omni}} to explore towards a more optimal weight. This optimization process is shown to be able to reach a higher performance in our study. 
Besides, compared with the adversarial training approach applied in our study, one advantage of \textbf{\textit{Omni}} is that it makes no assumptions of specific attacks, but explores the attack transferability property between the models.

Going forward, we would make a remark that it is recommended to try optimizing ensemble weights since this means that the contribution of each ensemble member to a prediction to be weighted proportionally to the trust of the performance of the member, which results in achieving better performance.

Another remark we would like to make is about the treatments in our study against potential adaptive attacks. In adaptive attacks~\cite{DBLP:conf/nips/TramerCBM20}, adversarial attackers design attacks specifically against a given defense mechanism. Adaptive attacks have attracted more attention for evaluating defenses to adversarial examples. A recent study from Tramer et al.~\cite{DBLP:conf/nips/TramerCBM20} reports that a diverse set of thirteen defense strategies can be circumvented with adaptive attacks. We make no claim that \textbf{\textit{Omni}} are robust to adaptive attack through this empirical study, however, we believe that it would be an interesting direction to explore for future work.

\subsection{Hyperparameter Distance}~\label{sec:distance}

\begin{figure}[!ht]
\centering
\includegraphics[width=9cm]{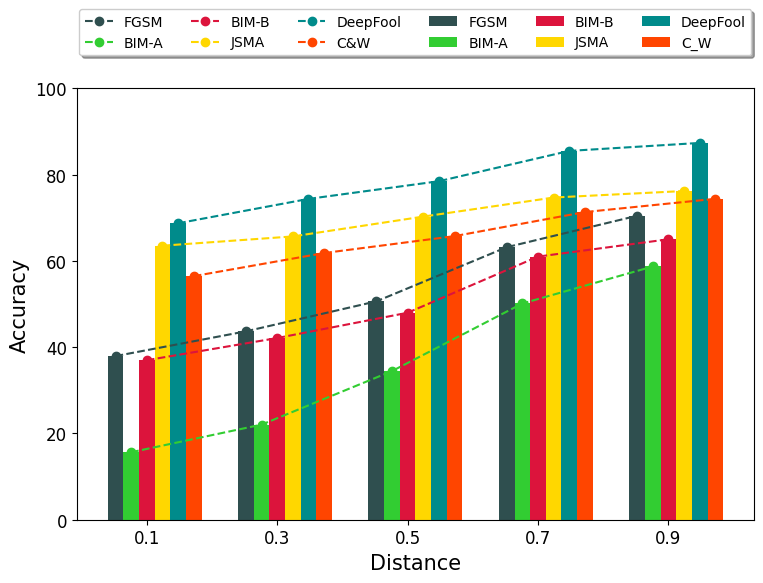}
\caption{Classification \textit{accuracy} (\%) of \textbf{\textit{Omni}} with different hyperparameter distance on dataset Contagio PDF.}
\label{fig:contagio_distance}
\end{figure}

\begin{figure}[!ht]
\centering
\includegraphics[width=9cm]{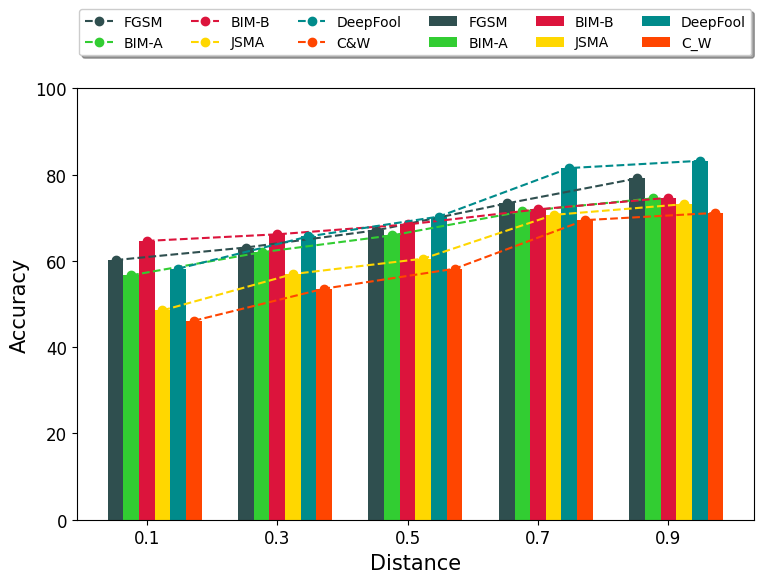}
\caption{Classification \textit{accuracy} (\%) of \textbf{\textit{Omni}} with different hyperparameter distance on dataset NSL-KDD.}
\label{fig:nsl-kdd_distance}
\end{figure}

\begin{figure}[!ht]
\centering
\includegraphics[width=9cm]{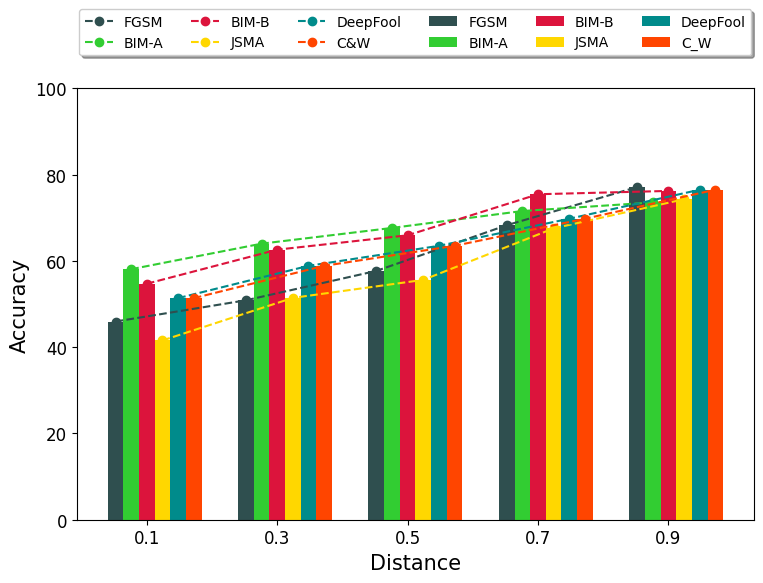}
\caption{Classification \textit{accuracy} (\%) of \textbf{\textit{Omni}} with different hyperparameter distance on dataset CIC-IDS-2017.}
\label{fig:cic-ids-2017_distance}
\end{figure}

\begin{figure}[!ht]
\centering
\includegraphics[width=9cm]{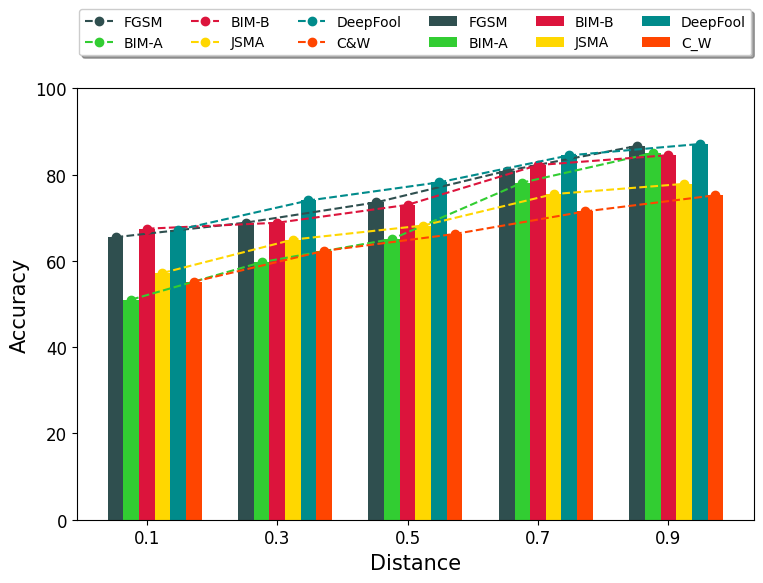}
\caption{Classification \textit{accuracy} (\%) of \textbf{\textit{Omni}} with different hyperparameter distance on dataset CSE-CIC-IDS2018.}
\label{fig:cse-cic-ids-2018_distance}
\end{figure}

\begin{figure}[!ht]
\centering
\includegraphics[width=9cm]{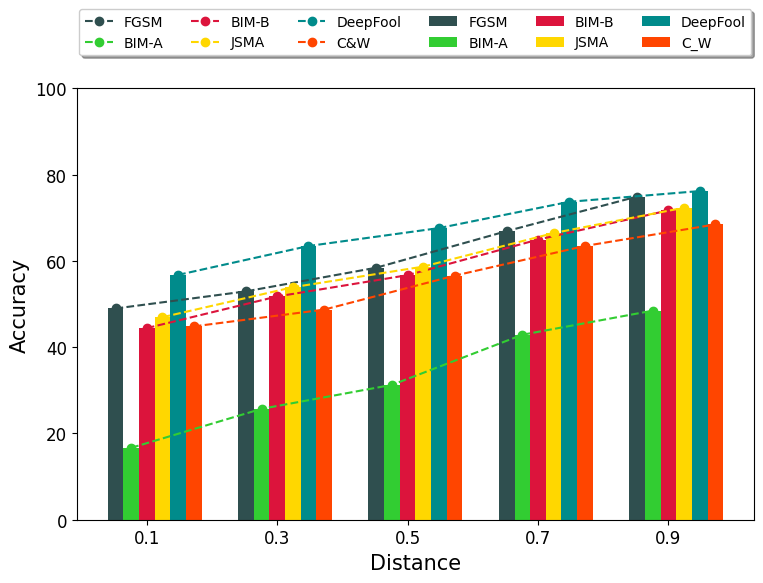}
\caption{Classification \textit{accuracy} (\%) of \textbf{\textit{Omni}} with different hyperparameter distance on dataset CICAndMal2017.}
\label{fig:cicandmal-2017_distance}
\end{figure}

In this section, we conduct a series of experiments to study the influence of the hyperparameter distance on the performance of \textbf{\textit{Omni}} against adversarial attacks. We only consider attacking \textbf{\textit{Omni}} in this case. We empirically experiment with different hyperparameter distances range from 0.1 to 0.9 (with a step of 0.2). This range explores ``unexpected models'' from a small distance to a large distance. The results are summarized in Figure~\ref{fig:contagio_distance} to Figure~\ref{fig:cicandmal-2017_distance}.

We observe that there is a trend shown (e.g., see the lines above the bar charts in Figure~\ref{fig:contagio_distance} to Figure~\ref{fig:cicandmal-2017_distance}) in the results table. A larger hyperparameter distance between the constitute models in the ensemble system and the ``victim model'' would help to build a defense that is less likely to be susceptible to attack transferability. Moreover, for \textbf{\textit{Omni}}, if $d$ is small, i.e., the architecture of selected models and the ``victim model'' are quite close, the defense performance is also close to $A_1$ results (i.e., attack accuracy). This trend provides useful suggestions for constructing stronger ensemble classifier in practice.

~\par

\section{Discussion}~\label{sec:discussion}

\subsection{The Nature of \textbf{\textit{Omni}}}

We now provide a brief discussion of why the proposed approach \textbf{\textit{Omni}} can help to build a more robust classifier. We make the following remarks as well as hypothesis from this empirical study. In the future, we plan to further validate this hypothesis empirically or theoretically.

\textit{Hyperparameter optimization.} In machine learning tasks, hyperparameter optimization is a widely-used technique that fine-tunes hyperparameters of models in order to achieve a better performance. An explanation of how hyperparameter optimization works is the change of model's decision boundary. For example, as for a machine learning algorithm such as SVM (Support Vector Machine) in a binary classification task, if the data points in different classes are linearly separable, then it is easy to draw a decision boundary. However, in some real cases, the noisy data points make it not trivial to separate the data linearly. A standard SVM would typically try to separate all data points and not to misclassify any point. This phenomenon usually results in an overfit model and even in some cases, a decision boundary cannot be found.

The idea of ``soft margin''~\cite{cortes1995support} in SVM allows some of the data points to be misclassified (i.e., on the wrong side of decision boundary), which helps to build a better generalized classifier. The ``soft margin'' tries to optimize a trade-off of two goals: 1) increase the distance of support vectors to decision boundary; 2) maximize the number of data points that are correctly classified. In the SVM algorithm, the \textit{C} parameter adds a regularization penalty for each misclassified data point and the \textit{gamma} parameter indicates the kernel trick. When optimizing the SVM-based models, the selection of different \textit{C} parameter and \textit{gamma} parameter would generate models with distinct decision boundaries.

We make one hypothesis about the decision boundary of models in \textbf{\textit{Omni}}. Close trained models (with similar inherent structure) share similar decision boundaries on the same datasets, while distant models (in terms of hyperparameter distance) would contribute to creating different decision boundaries. \textbf{\textit{Omni}} then select models with distinct decision boundaries from the target model. \textbf{\textit{Omni}} seeks ``decision boundary variance'' with these chosen models.


\textit{Ensemble classifier.} Compared to a single model, an ensemble classifier is supposed to make the attackers harder to figure out the decision boundary and thus more difficult to find adversarial examples. \textbf{\textit{Omni}}'s strategy is to build an ensemble classifier from models the decision boundaries of which are even distinct from the ``victim models''. A further exploration of \textbf{\textit{Omni}} would be increasing the diversity of ensemble. Prior study~\cite{DBLP:conf/icml/PangXDCZ19} demonstrates that a more diverse ensemble would increase the robustness than a less diverse ensemble.

\textit{Weight optimization.} The weighted ensemble is related to the ``voting ensemble'' which involves combination of predictions from multiple other models. A trivial trick is to average the prediction from each model. A limitation of this technique is that it assumes all models in the ensemble are equally effective, which may not be the case. \textbf{\textit{Omni}} further enforces another optimization effort in searching for optimal contribution of each constitute model.

\subsection{Trade Principle}

Results in Section~\ref{sec:result} show that the final $A_5$ accuracy is still less than the original $A_0$ pre-attack accuracy. Mathematically, we can show that this is the expected case:
\bi
\item
Khasawneh et al.~\cite{khasawneh2017rhmd} offer a mathematical analysis
of how hard it is for an adversary to reverse engineer the defense model.  
\item
Given an ensemble of $H$ learners, each of which has its own errors of $e(\hat{H_i})$, then using the probably approximately correct (PAC) learning theory~\cite{valiant84} we can show that the upper bound on the error of the attacker approximation of the defense model is 
\begin{equation}\label{one}
2(\mathit{max}\; e(\hat{H_i}))
\end{equation}
i.e. twice the worst error of any defense learner. This result has a clear intuition: the more the defense model makes mistakes, the harder it becomes for the attacker to learn the policies of the defense strategy. 
\item
Khasawneh et al.~\cite{khasawneh2017rhmd} warn that this kind of defense has an unwanted side-effect. Specifically, it can reduce prediction performance. We call this the \textit{trade principle}:
\begin{quote}
``{\em The above theorem (Equation~\ref{one}) suggests a trade-off between the accuracy of the defense model under no reverse-engineering vs. the susceptibility to being reverse-engineered: using low-accuracy but high-diversity classifiers allow the defender to induce a higher error rate on the attacker, but will also degrade the baseline performance against the target.}''
\end{quote}
\ei

The trade principle tells us that adversarial defense is mathematically required to lose predictive accuracy as they struggle to respond to an attack. That is, we should expect that $A_5$ is less than $A_0$ (the pre-attack accuracy). A system under attack suffers some predictive losses -- the issue is how much we can mitigate that loss.

\section{Threats to Validity}~\label{sec:threats}

As to any empirical study, biases can affect the final
results. Therefore, conclusions drawn from this work must be
considered with threats to validity in mind. In this section, we
discuss the validity of our work.

\textit{\textbf{Attack strategy bias.}} We evaluate our method over a diverse set of attack strategies. However, we do not argue that our list of attacks is exhaustive, and we observe that more or more complicated adversarial attacks are emerging in recent work, which can be one of our future research directions.

\textit{\textbf{Sampling bias.}} Our datasets cover network traffic, Android malwares and malicious pdf files, which are representative in previous intrusion detection and malware detection related research. However, some other popular data in existing work are also available which be further evaluated.

\textit{\textbf{Evaluation bias.}} Carlini et al.~\cite{carlini2019evaluating} introduce several commonly accepted best practices that can be used to evaluate the defenses to adversarial examples. One of the suggestions is to protect from the adaptive adversaries, which means they are adapted to the specific details of the defense and hence further invalidate the robustness. Our approach is not designed with adaptive in mind, which is one of our future work.

\textit{\textbf{Parameter bias.}} There are a bunch of parameters that control the degree of perturbation to the dataset. For example, a larger epsilon value cause large perturbation which further reduces the accuracy under attack. We do explore every set of parameters or the converge issue as suggests by Carlini et al.~\cite{carlini2019evaluating}. In addition, Table~\ref{tbl:hyperparameters} mentions the hyperparameter options explored but the hyperparameter optimization spaces are much more vast. Exploring these options will require vast amount of CPU resources. Thus we will require to make a trade-off between exploring the extend of hyperparameter space and cost awareness of the models. We do not claim that hyperparameters that we select for Bayesian optimization are exhaustive, rather, we believe the hyperparameter ranges that we choose are enough for us in the work.

\section{Conclusion}~\label{sec:conclusion}

When attackers can fool a classifier to think that a malicious input is actually benign, they can render a machine learning-based malware or intrusion detection system ineffective. Various researchers have proposed ensemble methods as a technique to increase the complexity and robustness of a defense model~\cite{DBLP:conf/iclr/TramerKPGBM18,smutz2016tree,kantchelian2016evasion}. In theory, this added complexity makes it harder for the attackers to learn the defender's model. However, several researchers have reported that such ensemble learners have limited advantages.
 
The starting point for  this paper was the observation that prior
work barely scratched the surface of all the options available when building an ensemble system. Rather than use a handful of models (as done by, eg.   Kantchelian et al.~\cite{kantchelian2016evasion}), we explore a much large space of options within the hyperparameter configuration space of a model.  
 
In our approach, we  create an ensemble of ``unexpected models''; i.e., models whose control hyperparameters have a large distance to the  hyperparameters of an adversary's target model. In studies with five   adversarial evasion attacks   on five security datasets, we show that this method can successfully mitigate the effects of adversarial evasion attack.    Hence we conclude:
\begin{quote}
{\em
A well-designed weighted ensemble system is a promising approach to defend against adversarial evasion attack.
}\end{quote}
and
\begin{quote}
{\em
When using ensemble learning as a defense method against adversarial evasion attacks, we suggest to create ensemble with unexpected models that are distant from the attacker's expected model (i.e., target model) through methods such as hyperparameter optimization.
}\end{quote}

For future directions, we recommend trying to
speed up our optimization methods.
Table~\ref{tbl:optimizationTime} shows the runtime of the optimization process: TPE takes around nine hours to terminate (on average). Clearly, this needs to be improved. When attackers have more knowledge and resources than defenders, they might be able to learn to adapt faster than we can defend against. Therefore, it is vital that we make our method as fast as possible. 

\begin{table}[!htbp]
\centering
\caption{Runtime of Bayesian optimization.}
{\footnotesize\begin{tabular}{l|c}
\hline
\rowcolor[HTML]{ECF4FF} 
Dataset
&   Runtime H:M:S  \\ \hline
CIC-IDS-2017 & 10:33:31 \\ 
CICAndMal2017 & 9:22:58  \\ 
ContagioPDF & 1:47:29 \\ 
CSE-CIC-IDS2018 & 15:59:02 \\ 
NSL-KDD & 6:54:22 \\ \hline
\end{tabular}}
\label{tbl:optimizationTime}
\end{table}

Also, here we were only optimizing for accuracy. But Carlini et. al~\cite{carlini2019evaluating} suggest to that other metrics might be equally important such as TP (true positives), TN (true negatives), FP (false positives) and FN (false negatives). Here we have not explored such multi-goal reasoning (since that would be much slower) but this is clearly a direction for future work.
    
Finally, here  we have assumed that the attack strategies are constant across our experiments. In the future work, it would be also important to explore a more challenging {\em adaptive adversaries} that frequently change their attacks strategies.

\begin{acknowledgements}
This work was partially funded by NSF grant \#1909516.
\end{acknowledgements}

\bibliographystyle{spbasic} 
\bibliography{main.bbl}






\end{document}